\documentclass[%
 reprint,
 amsmath,amssymb,
 aps,
]{revtex4-2}

\usepackage{graphicx}
\usepackage{dcolumn}
\usepackage{bm}
\usepackage{amsmath}
\usepackage{braket}
\usepackage{xcolor}

\usepackage[utf8]{inputenc}

\begin{document}

\preprint{APS/123-QED}

\title{Polarization Engineering of the Orbital Hall Conductivity in Two-dimensional Ferroelectric Higher-Order Topological Insulator Tl$_2$S and SnS}

\author{Yingjie Hu$^1$}
\author{Heng Gao$^1$}
 \email{gaoheng@shu.edu.cn}
\author{Yabei Wu$^2$}
 \email{wuyb3@sustech.edu.cn}
\author{Wei Ren$^1$}
 \email{renwei@shu.edu.cn}

\affiliation{$^1$Physics Department, Materials Genome Institute, Institute for Quantum Science and Technology, Shanghai Key Laboratory of High Temperature Superconductors, Shanghai University, Shanghai 200444, China
}%

\affiliation{$^2$Department of Materials Science and Engineering and Guangdong Provincial Key Lab for Computational Science and Materials Design, Southern University of Science and Technology, Shenzhen, Guangdong 518055, China}%


\begin{abstract}
Ferroelectric higher-order topological insulators (HOTIs) exhibit tunable physical properties arising from the interplay between ferroelectric polarization and band topology. This work investigates the topological origin of two classes of two-dimensional (2D) ferroelectric HOTIs with out-of-plane or in-plane polarization, revealing their distinct orbital transport behaviors and the mechanism for engineering orbital Hall conductivity (OHC) via polarization control. Our results demonstrate the unique role of polarization in modulating both the higher-order band topology and orbital transport. A strong coupling between in-plane polarization and higher-order topology is identified, establishing in-plane polarization as an intrinsic means to reversibly switch the OHC plateau within the band gap. Using Tl$_2$S and SnS as representative models of the two HOTI types, we demonstrate persistent and electrically switchable orbital transport, respectively. Our study advances the understanding of the coupling among ferroelectricity, higher-order topology, and orbital transport, offering new avenues for controllable orbitronics.
\end{abstract}

\maketitle


\section{Introduction}
Topological insulators (TIs) represent a distinct class of quantum materials characterized by robust, topologically protected gapless states at their edges or surfaces~\cite{W1,W2,W3}. These exotic boundary states lead to various quantized electromagnetic phenomena, including the quantum spin Hall effect and quantized magnetoelectric responses~\cite{W4,W5}. Recently, the concept of ferroelectric higher-order topological insulators (HOTIs) has further extended the principle of bulk-boundary correspondence, where nontrivial topological features manifest not on conventional ($d-1$)-dimensional boundaries but on lower-dimensional ones~\cite{W6,W7}. Specifically, an $m$-dimensional $n$th-order topological insulator is defined by the presence of gapless states at its ($m-n$)-dimensional boundaries. Unlike conventional TIs, whose topological edge states arise from spin-orbit coupling (SOC), higher-order topological phases are generally protected by crystalline symmetries—such as rotation or inversion symmetry~\cite{W8}. To date, HOTIs have been theoretically predicted in several material systems, including transition metal dichalcogenides (TMDs)~\cite{W9,W10,W11,W12}, twisted bilayer graphene~\cite{W13,W14}, and certain magnetic materials~\cite{W57,W58,W59,W60}. Experimentally, hinge states associated with HOTIs have been observed in three-dimensional materials such as bismuth (Bi)~\cite{W15,W16} and tungsten ditelluride (WTe$_2$)~\cite{W17,W18,W19}.

In two dimensions, spin Chern insulators (SCIs) host the quantum spin Hall effect and support helical edge states. Within the band gap, a quantized plateau emerges in the spin Hall conductivity (SHC), which is theoretically exact in the ideal case. However, small deviations from exact quantization can arise due to the non-conservation of spin angular momentum. For HOTIs, Costa \textit{et al.} identified a nonzero SHC plateau inside the band gap, serving as a signature of the nontrivial topology. Their eight-band tight-binding model predicts that this SHC plateau takes finite values between $-e^2/h$ and $+e^2/h$. Similarly, the orbital Hall effect (OHE) describes the transverse flow of orbital angular momentum (OAM) induced by a longitudinal electric field. Unlike the spin Hall effect, the OHE does not necessarily require strong SOC. Earlier theoretical studies have predicted that orbital textures in two-dimensional (2D) materials can generate a pronounced OHE, and this effect has been experimentally observed in TMDs with weak SOC. By analogy with the quantized SHC plateau in HOTIs, Costa \textit{et al.} linked the emergence of a HOTI phase to an orbital Hall insulating phase in 1T-structured TMDs. Correspondingly, a nonzero OHC plateau is expected to appear within the band gap, reflecting characteristic orbital transport in HOTIs. Compared with spin, the eigenstates of orbital angular momentum act as pseudospins and are simultaneously eigenstates of the $C_3$ rotation operator that protects the HOTI phase. Therefore, the higher-order topological phase can also be detected through measurements of the OHE.

In this work, we employ density functional theory (DFT) to study the higher-order topological properties and polarization-regulated orbital transport in 2D HOTIs with out-of-plane or in-plane polarization through specific material examples. In out-of-plane ferroelectric systems, polarization generally coexists with but does not directly drive changes in higher-order topology. We reveal the symmetry-protected origin of the higher-order topology in such systems.

Using Tl$_2$S as a representative example, we identify robust rotation-symmetry-protected higher-order topological corner states and demonstrate persistent orbital transport that remains unaffected by polarization switching. In contrast, in-plane polarization can induce higher-order topological phase transitions in ferroelectric systems. As a result, the orbital Hall conductivity (OHC) plateau within the band gap—a hallmark of HOTIs—shifts from zero to a finite value during these transitions. Taking SnS as a material candidate, we uncover the strong coupling between in-plane polarization and higher-order topology and show that orbital transport can be effectively controlled via polarization reorientation. Furthermore, we calculate the orbital Berry curvature to elucidate the physical mechanism underlying the polarization-mediated regulation of orbital Hall conductivity.

\begin{figure}[t]
  \includegraphics[width=0.45\textwidth]{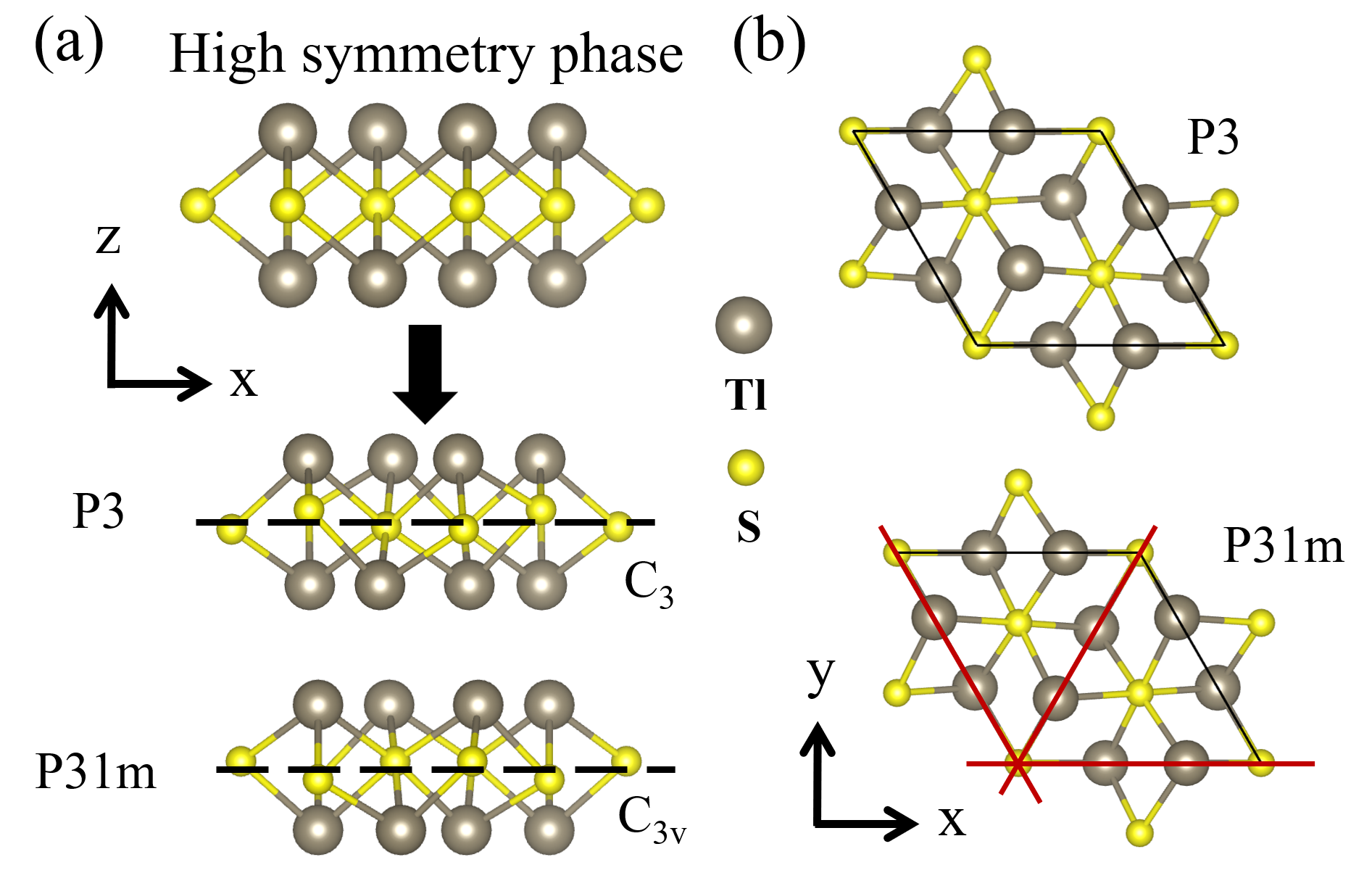}
  \caption{\label{fig1}(a) Crystal structure of Tl$_2$S in its high-symmetry phase (space group P3m1) and the two corresponding ferroelectric phases (space groups P3 and P31m, respectively). Brown and yellow spheres represent Tl and S atoms. 
(b) Structural differences between the P3 and P31m phases. Red lines indicate the mirror symmetry axes.}
\end{figure}

\section{High-order topologies in out-of-plane ferroelectric systems}

In this section, we focus on the higher-order topological origin and the distinct orbital transport properties in out-of-plane ferroelectric systems. As an emerging 2D ferroelectric material, Tl$_2$S has attracted considerable research interest due to its diverse ferroelectric and antiferroelectric phases. Using two of its ferroelectric phases as examples, we clearly demonstrate the robust higher-order topological corner states protected by rotational symmetry. Furthermore, by studying the transition states between ferroelectric phases, we reveal a stable OHC plateau within the band gap and elucidate its origin through orbital Berry curvature analysis.

Tl$_2$S crystallizes in an anti-TMD structure, where S atoms occupy the metal sites typically held by transition metals in conventional TMDs, while Tl atoms reside at the chalcogen positions. This distorted anti-TMD configuration was previously identified in early experimental studies (see Refs.~\cite{W29,W30,W31}). Its high-symmetry phase adopts a 1T structure with space group  $P\bar{3}m1$, as illustrated in upper panel of Fig.~\ref{fig1}a. 
 
Similar to conventional TMD materials such as MoS$_2$, the 1T phase of Tl$_2$S is energetically unstable. Previous studies (e.g., Ref.~\cite{W32}) report three low-energy polymorphs: two ferroelectric phases and one antiferroelectric phase. In this work, we focus exclusively on the two ferroelectric phases of Tl$_2$S, which belong to the space groups P31m and P3, respectively. Both structures break inversion symmetry, giving rise to out-of-plane ferroelectric polarization. Although the P3 and P31m phases differ only slightly in atomic positions and lattice parameters, they exhibit distinct symmetry characteristics: both possess $C_3$ rotational symmetry, but the 
P31m phase additionally retains three mirror symmetries, indicated by red lines in Fig.~\ref{fig1}. These structural differences arise from distinct lattice vibrational modes, leading to different magnitudes of out-of-plane dipole moments. In the following, we systematically investigate the higher-order topological properties of these two ferroelectric structures.

\begin{figure*}[htbp]
  \includegraphics[width=0.96\textwidth]{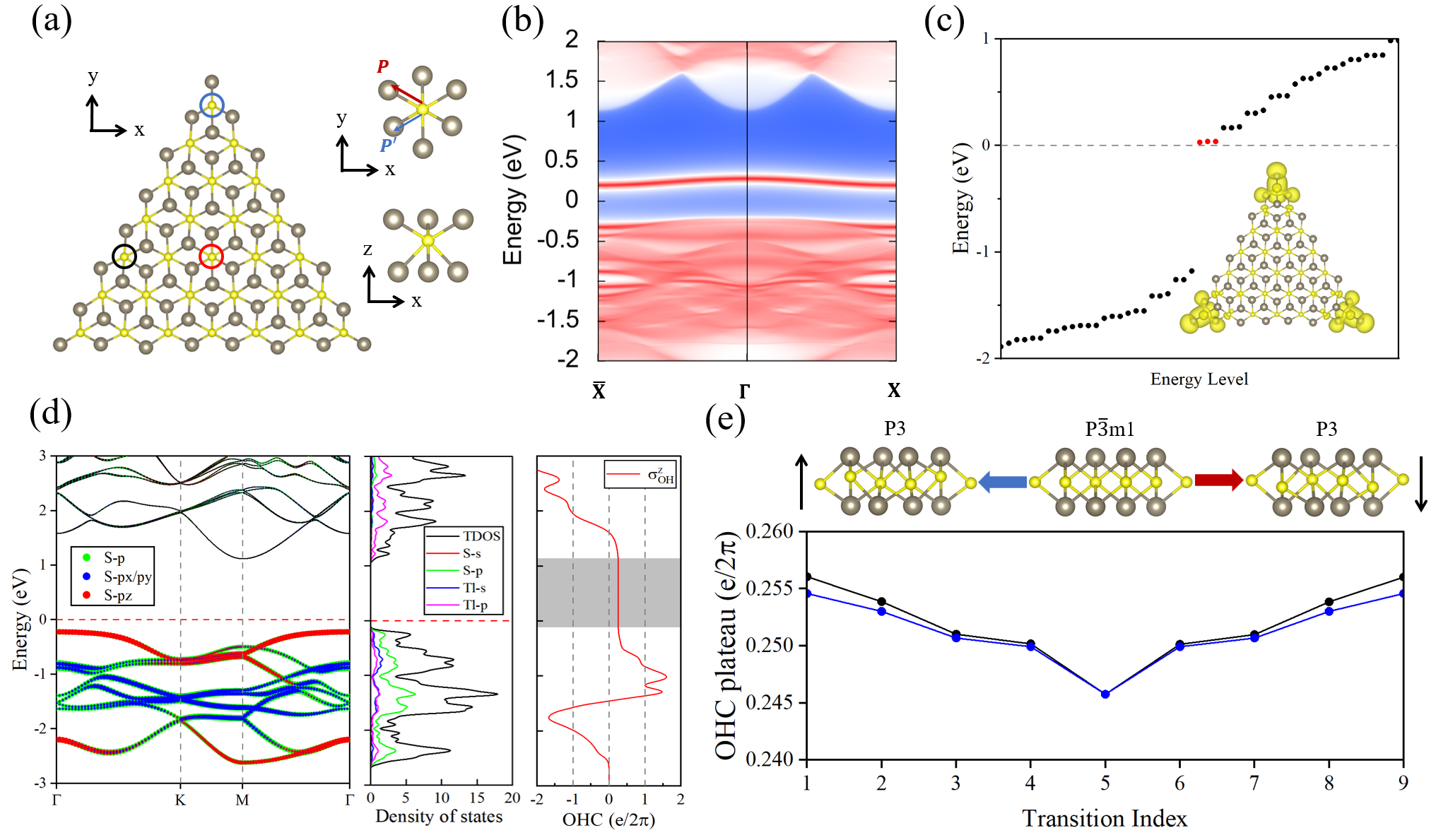}
  \caption{\label{fig2}
  (a) Finite triangular nanoflake of the P3 phase (top and side views of the Tl$_6$S octahedron). The red, black, and blue circles correspond to atoms in bulk, edge, and corner environments, respectively. (b) Edge states of the P3-phase Tl$_2$S nanoribbon. A floating edge state appears between the valence and conduction bands. (c) Energy spectrum of the P3-phase Tl$_2$S triangular nanoflake. Red dots indicate corner states, with their real-space distribution shown in the corresponding panel. (d) Electronic structure and orbital transport of the P3 phase. Shown are the band structure, density of states, and orbital Hall conductivity (OHC) as a function of Fermi energy position. (e) Ferroelectric switching in the P3 phase and associated OHC plateaus. The ferroelectric switching path (index 1 to 9, with 5 denoting the intermediate phase) and the corresponding OHC plateaus within the band gap are displayed. The blue curve shows the switching process for the P31m phase for comparison.}
\end{figure*}

Taking the P3 phase of Tl$_2$S as an example, we theoretically illustrate the emergence of higher-order topological corner states protected by rotational symmetry, as well as the out-of-plane polarization resulting from broken inversion symmetry. In a finite triangular nanoflake, we analyze each Tl$_6$S octahedron in the bulk [Fig.~\ref{fig2}(a)]. Within these octahedra, the S–Tl bonds preserve $C_3$ rotational symmetry, which suppresses the formation of in-plane polarization. At edges and corners, however, the cancellation of in-plane polarization is incomplete, driving electron accumulation and leading to the formation of edge and corner states. In contrast, out-of-plane polarization originates from the incomplete compensation of Tl–S bond distortions along the z-direction due to the loss of inversion symmetry. Consequently, while the higher-order topological corner states are protected by rotational symmetry in a finite nanoflake, out-of-plane polarization exists throughout the entire structure without breaking the rotational symmetry. Using DFT calculations, we confirm the higher-order topological nature in both the P3 and P31m phases. Furthermore, during out-of-plane polarization switching, these higher-order topological features remain robust.
\begin{eqnarray}
\label{4}
  P_{bulk}^{in-plane}&=\sum_i\left[(P_i+C_3P_i+C^{-1}_3P_i) \right.\\ \nonumber
  &\left.+(P'_i+C_3P'_i+C^{-1}_3P'_i)\right]=0.
\end{eqnarray}

The crystal structure of the $P3$ phase is shown in Fig.~\ref{fig1}(b), that possesses only $C_3$ rotation symmetry. Phonon spectrum calculations [Fig. S3(b)] show no imgaginary frequencies, confirming its dynamic stability. Electronic band structure analysis reveals that the P3 phase is a semiconductor with a band gap of 1.34 eV [Fig.~\ref{fig2}(d)]. Density of states (DOS) and fatband analysis indicate that the top of the valence band is primarily composed of S-$p$ orbitals, while the bottom of conduction band is dominated by Tl-$p$ orbitals. 

To further investigate its topological nature, we construct a tight-binding (TB) model using the maximally localized Wannier functions (MLWFs) based on the $p$ orbitals of Tl and S. The calculated edge states of a nanoribbon [Fig.~\ref{fig2}(b)] exhibit no gapless states connecting the valence and conduction bands, initially suggesting a topologically trivial insulator in the first-order sense. However, the emergence of floating edge states serves as a distinct signature of higher-order topological phases.

The nontrivial topological classes arising from the $C_n$ symmetry can be distinguished by the symmetry representations at the high symmetry points (HSPs) of the occupied bands~\cite{W33}. Following the definition in Ref.{\cite{W33}}, we denote the HSPs as $\Pi$ and the eigenvalues of the $C_n$ rotation operator at HSP $\Pi$ as $\Pi_p^{(n)}=e^{i2\pi(p-1)/n}$ ($p=1,2,...,n$). If the eigenvalues change at different HSPs, the nontrivial topology arises. The topological invariants are defined relative to $\Gamma=(0,0)$ as,
\begin{equation}
  \label{1}
  \left[\Pi_p^{(n)}\right]=\#\Pi_p^{(n)}-\#\Gamma_p^{(n)},
\end{equation}
where $\#\Pi_p^{(n)}$ is the number of occupied bands with the eigenvalue $\Pi_p^{(n)}$. However, not all the invariants are independent. For the ferroelectric phase $P3$ and $P31m$, both of which lack inversion symmetry but retain $C_3$ symmetry, the topological indices are given by: 
\begin{eqnarray}
  \label{2}
\chi^{(3)}=\left(\left[K_1^{(3)}\right],\left[K_2^{(3)}\right]\right),
  \\
  \label{3}
  Q_{\rm{corner}}^{(3)}=\frac{e}{3}\left[K_2^{(3)}\right] \rm{mod}\ e.
\end{eqnarray}
Apply the above topological index method, firstly we calculate the symmetry eigenvalues of the occupied bands of $P3$-Tl$_2$S. By calculating its band irreducible representation~\cite{W34}, we get a nonzero topolagical index of [-2,1], suggesting that $P3$-Tl$_2$S is a potential HOTI with a fractional corner charge $Q_{\rm{corner}}^{(3)}=e/3$. To reveal the corner states, we construct a finite nanoflake and calculate its energy spectrum at the single Gamma point. The energy spectrum and the corresponding charge density distribution in real space is shown in Fig.~\ref{fig2}(c). Near the Fermi level, there exists degenerate energy levels which can be recognized as corner states by the real-space distribution. The charge density is uniformly distributed at the three corners and the charge at each corner should be $e/3$. 

The mismatch between the number of electrons in energy band and the number of electrons required for charge neutrality causes filling anomaly~\cite{W35,W36}. This filling anomaly, accompanied with the introduction of degenerate corner states, giving consequence to the fractional quantization of corner charge.

For the P31m phase, another polar phase of Tl$_2$S, the crystal structure is presented in Fig.~\ref{fig1}(b). This phase possesses three additional mirror symmetry operations compared to the $P3$ phase. Phonon calculations confirm its stability, as evidenced by the absence of negative frequencies [Fig.~S3(a)]. Although the $P31m$ phase exhibits a distinct distortion pattern and correspondingly different structural and polarization characteristics from the $P3$ phase, it retains the $C_3$ rotational symmetry that protects higher-order topological corner states. Consequently, our analysis identifies the $P31m$ phase as a higher-order topological insulator (HOTI) as well. Similar to the $P3$ phase, all relevant higher-order topological features of this structure are systematically analyzed in the Supplemental Material.

\subsection{Orbital Hall conductivity of Ferroelectric HOTI Tl$_2$S}
As an emerging field of orbitronics, orbital Hall effect has attracted extensive attention. Recently, plenty of works have studied the system with non-zero OHC plateaus within their insulating band gaps that are called orbital Chern insulators~\cite{W38,W39,W40,W41}, and it is possible to attribute an orbital Chern number to this insulating phase~\cite{W28}. Here, we mainly focus on the connection between the orbital Hall insulating phase and HOTIs. From the linear response theory, the OHC can be calculated by~\cite{W55}
\begin{equation}
  \label{5}
  \sigma _{xy}^{\hat{L}_z}=\frac{e\hbar}{(2\pi)^2}\sum_{n}\int_{\rm{BZ}}d^2kf_{n,k}\Omega _{n,xy}^{\hat{L}_z}(k),
\end{equation}
where $f_{nk}$ is the Fermi-Dirac distribution and $\Omega _{n,xy}^{\hat{L}_z}(k)$ is the orbital-weighted Berry curvature,
\begin{equation}
  \label{6}
  \Omega _{n,xy}^{\hat{L}_z}(k)=2 \sum_{m\ne n}\mathrm{Im} \left[\frac{\bra{\psi_{n,k}}j_{y}^{\hat{L}_z}\ket{\psi_{m,k}}\bra{\psi_{m,k}}\hat{v}_x\ket{\psi_{n,k}}}{(E_{n,k}-E_{m,k})^2}\right].
\end{equation}
Here $\ket{\psi_{n,k}}$ and $E_{n,k}$ denote Bloch eigenstates and eigenvalues of the Hamiltonian in reciprocal space. The velocity operator is $v_i=\frac{1}{\hbar}\frac{\partial H(k)}{\partial k_i}(i=x,y)$ and the orbital angular momentum (OAM) current operator component along $y$ with polarization $z$ is defined as $j_{y}^{\hat{L}_z}=\left[\hat{L}_zv_y+v_y\hat{L}_z \right]/2$.

Figure~\ref{fig2}(d) presents the band structure, DOS, and OHC of the $P3$-Tl$_2$S. A finite OHC plateau is clearly observed within the band gap, which, according to the Ref {~\cite{W28}}, serves as a characteristic signature of HOTIs. Thus, the OHC plateau in Tl$_2$S is consistent with the emergence of higher-order topological corner states. 

However, compared to the other known HOTIs, the magnitude of the OHC plateau in Tl$_2$S is relatively small. This can be understood by examining the orbital composition and band topology. The DOS indicates that both the top of the valence band and the bottom of conduction band are primarily derived from the $p$-orbitals of S and Tl. According to Eq.~(\ref{5}), the OHC is mainly composed of the valence band near the Fermi level which is $p$ orbital of S. Therefore, the OAM $\hat{L}_z$ would not have a large expectation value. On the other hand, the gap is indirect such that the $\Omega _{n,k}$ has no strong peak at a certain point, being more spread throughout the Brillouin zone (BZ). We calculate the orbital-weighted Berry curvature to illustrate the OHC plateau, as shown in Fig.~\ref{fig3}(c).

\begin{figure}[htbp]
  \includegraphics[width=0.48\textwidth]{Fig3.png}
  \caption{\label{fig3} Orbital-weighted Berry curvature of intermediate phase Tl$_2$S (a,b) and ferroelectric-phase Tl$_2$S (c,d), plotted along high-symmetry paths (a,c) and across the full Brillouin zone (b,d).}
\end{figure}

In ferroelectric materials, the polarization switching necessarily involves an intermediate phase~\cite{W42}. For Tl$_2$S, the intermediate phase is the high-symmetry phase $P\bar{3} m1$. It is completely consistent with the 1T phase TMDs which have already been proved as HOTIs~\cite{W28,W43}. The higher-order topological properties of high-symmetry phase Tl$_2$S are also protected by the inversion symmetry. Based on the polar $P3$ phase and high-symmetry phase $P\bar{3} m1$, we investigated the ferroelectric switching process. In this process, although the inversion symmetry is broken, the $C_3$ rotation symmetry remains to protect the nontrivial corner states. Therefore, the ferroelectric polarization transition is continuous in a topological sense. So we make a nudged elastic band (NEB) calculation~\cite{W45}. By linear interpolation, we obtain a series of transition states which are all theoretically HOTIs. Then we calculate the OHC of each structure, as shown in Fig.~\ref{fig2}(e). 
We find that the OHC plateaus remain continuous and finite, indicating the absence of a higher-order topological phase transition. Consistent results are observed in the P31m ferroelectric phase, where higher-order topological features are likewise robust against out-of-plane polarization switching, as shown by the blue curve in Fig.~\ref{fig3}(e). Additionally, we calculate the orbital-weighted Berry curvature for the intermediate phase [Fig.~\ref{fig3}(a,b)], which reveals only minor differences compared to the polar phases.


\section{High-order topologies and polarization tunable orbitronics in in-plane ferroelectric systems}

In the Tl$_2$S system discussed above, the robustness of higher-order topological properties against polarization switching leads to persistent orbital transport throughout the ferroelectric reversal process. In contrast, for in-plane ferroelectric systems, polarization can couple directly with higher-order topology due to crystallographic symmetry. Under an in-plane electric field, electrons gain a transverse velocity and give rise to corner states, as suggested in previous studies (e.g., Refs. \cite{W37,W56}). Consequently, introducing in-plane polarization in such systems can drive higher-order topological phase transitions. As a hallmark of HOTIs, distinct orbital transport signatures emerge accordingly. In the candidate material SnS, we demonstrate a reversible switching of the OHC plateau via a polarization-induced higher-order topological phase transition.

\begin{figure*}[htbp]
  \includegraphics[width=0.85\textwidth]{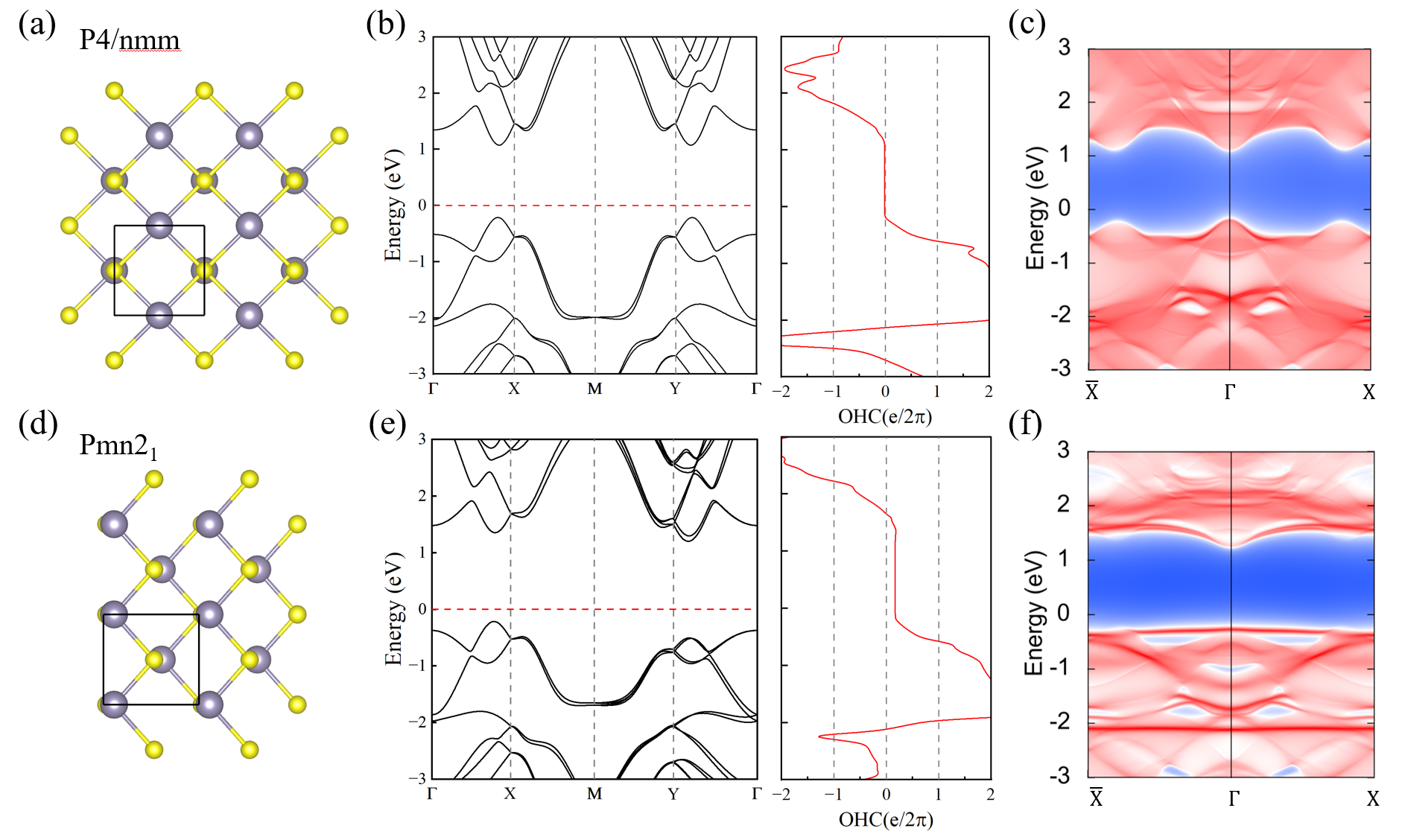}
  \caption{\label{fig4} (a) The crystal structure of intermediate phase $P4/nmm$-SnS. (b) The band structure and the variation of OHC with respect to the position of Fermi energy for $P4/nmm$-SnS. (c) The edge state of $P4/nmm$-SnS. (d) The crystal structure of ferroelectric phase $Pmn2_1$-SnS. The inversion symmetry and $C_{2y}$ rotation symmetry is broken. (e) The band structure and the variation of OHC with respect to the position of Fermi energy for $Pmn2_1$-SnS. (f) The edge state of ferroelectric phase $Pmn2_1$ with the floating edge states.}
\end{figure*}

We examine the established 2D ferroelectric HOTI SnS, as reported in Ref.~\cite{W37}. Orthorhombic SnS is an intrinsic multiferroic material exhibiting in-plane polarization~\cite{W46,W47,W48}. Its intermediate phase belongs to the space group $P4/nmm$, while the ferroelectric phase crystallizes in the polar space group $Pmn2_1$.

The introduction of polarization breaks the rotation and inversion symmetry. Specifically, in the absence of polarizaiton, the square lattice has both $C_{2x}$ and $C_{2y}$. When the polarizaiton is induced, for example, perpendicular to the $y$ direction, $C_{2y}$ rotation symmetry is broken, while $C_{2x}$ rotation symmetry survives and ensures the existence of corner states. Hence, in the SnS system, higher-order topological corner states are absent in the nonpolar phase but emerge in the polar phase~\cite{W37}.

\begin{figure}[h]
  \includegraphics[width=0.48\textwidth]{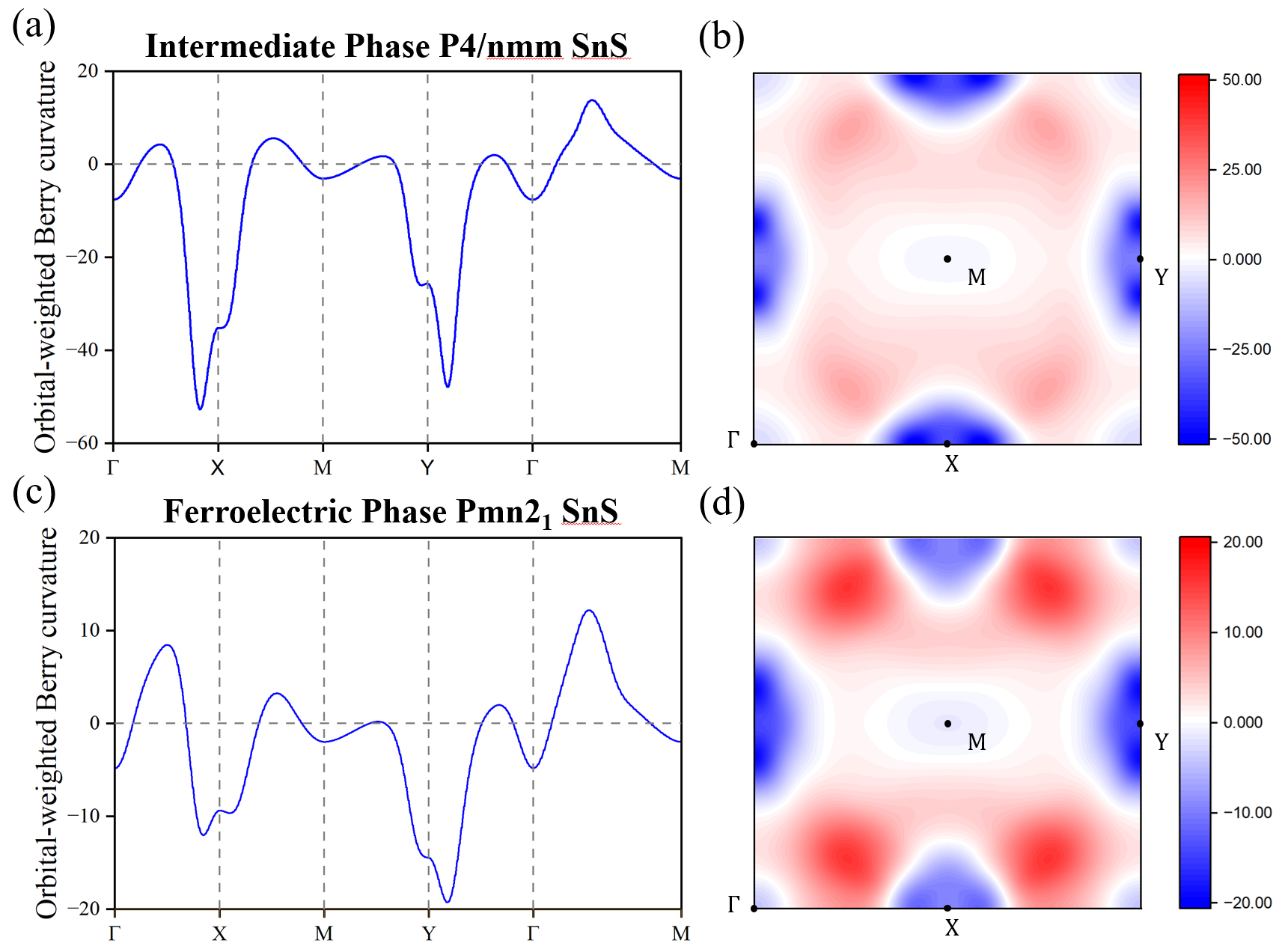}
  \caption{\label{fig5} Orbital-weighted Berry curvature of intermediate phase $P4/nmm$-SnS (a, b) and ferroelectric-phase $Pmn2_1$-SnS (c, d), plotted along high-symmetry paths (a, c) and across the full Brillouin zone (b, d).}
\end{figure}

The electronic structure and energy bands of the ferroelectric phase and the intermediate phase SnS are shown in the Fig.~\ref{fig4}(a,b,e,f). The broken of C$_4$ rotational symmetry leads to significant differences in the energy bands along the M-Y-$\Gamma$ path. It can be known through the DOS of ferroelectric phase which is shown in Fig.~S3(a,b) that the energy band near the Fermi level is mainly contributed by the $p$ orbitals of Sn and S, same as the intermediate phase. We construct the tight binding (TB) models using the $p$ orbitals of both Sn and S. The OHC of two phases calculated by the TB-model is shown in Fig.~\ref{fig4}(c,f). The results reveal that for high-order topological phase SnS, there exists a finite OHC plateau within the band gap, while for intermediate non-polarized phase SnS, the OHC plateau is zero. It can be known that the OHC in band gap can be served as a signature for identifying HOTIs, hence our calculation of the OHC is consistent with its high-order topological phase transition. Our results indicate that there is a OHC plateau switching induced by in-plane polarization accompanied by the phase transition. To further illustrate this, we calculate the edge states of the intermediate and higher-order topological phases, as shown in Fig.~\ref{fig4}(g,h). In higher-order topological phase, the edge states have an additional floating edge state relative to the intermediate phase. The floating edge state carries orbital angular momentum, which corresponds to the nonzero OHC plateau within the band gap. Additionally, we calculate the orbital-weighted Berry curvature to illustrate the switching behavior of the OHC plateaus, as shown in Fig.~\ref{fig5}. For the intermediate phase, the integral of the orbital-weighted Berry curvature over the entire Brillouin zone is zero due to complete cancellation between positive and negative contributions. In contrast, in the ferroelectric phase, the presence of polarization disrupts this compensation and results in a finite integrated value.

In summary, the introduction of in-plane polarization induces its high-order topological phase transition, and this phase transition is accompanied by the switch of the OHC plateau within the band gap from zero to finite. Therefore, in this type of materials where in-plane polarization induces high-order topological phase transitions, we propose to achieve the polarization regulation of the OHC plateau.

\section{Computational Methods}
The electronic structure calculations were performed by using density functional theory (DFT) and the projector augmented wave (PAW) method~\cite{W49}, as implemented in the Vienna Ab initio Simulation Package (VASP)~\cite{W50}. The exchange-correlation interactions were treated within the generalized gradient approximation (GGA) using the Perdew-Burke-Ernzerhof (PBE) functional. The energy cutoff for the plane-wave basis set is 500 eV, and all structures are relaxed until the energy and the forces are converged to 10$^{-8}$ eV and 0.001 eV/Å, respectively. The Brillouin zone was sampled using a 9$\times$9$\times$1  Monkhorst-Pack k-point mesh. A vacuum layer of 15 Å is used to avoid interactions between the nearest neighboring slabs. The maximally localized  Wannier functions are constructed using the WANNIER90 code~\cite{W51} and the edge state calculations were done using WannierTools~\cite{W52}. The irreducible representations were performed using IRVSP~\cite{W34} and the phonon dispersion is calculated with PHONOPY package~\cite{W53}. Moreover, the orbital Hall conductivity calculations are performed using our modified version of the open-source software WannierTools.

\section{Conclusions}
In conclusion, we have systematically investigated the higher-order topological origin and the tunability of the OHC plateau within band gaps across two distinct classes of ferroelectric systems. For out-of-plane ferroelectrics, we demonstrate that symmetry-protected higher-order topology coexists with out-of-plane polarization in the previously unreported HOTI Tl$_2$S. Throughout the ferroelectric switching process in Tl$_2$S, the polarization does not influence the higher-order topological properties, and consequently cannot modulate the OHC. In contrast, in in-plane ferroelectric systems, polarization couples with higher-order topology through crystallographic symmetry, enabling polarization-driven higher-order topological phase transitions. During these transitions, the OHC plateau within the band gap—a hallmark of HOTIs—shifts from zero to a finite value. Using SnS as a prototype system, we have successfully demonstrated polarization-mediated modulation of the OHC plateau inside the band gap.

\begin{acknowledgments}
This work was supported by National Natural Science Foundation of China (Grants No. 52130204, No. 12311530675, No. 12204299), Science and Technology Commission of Shanghai Municipality (Grants No. 22XD1400900), Shanghai Technical Service Center of Science and Engineering Computing, Shanghai University.
\end{acknowledgments}

\bibliography{apssamp}

\end{document}